# Polarization-dependent and Valley-protected Lamb Waves in Asymmetric Pillared Phononic Crystals

Wei Wang[1], Bernard Bonello[1*], Bahram Djafari-Rouhani[2], and Yan Pennec[2]

[1]Sorbonne Université, UPMC Université Paris 06 (INSP–UMR CNRS 7588),
4, place Jussieu 75005 Paris, France

[2]Institut d'Electronique, de Micro-électronique et de Nanotechnologie (IEMN–UMR CNRS 8520),
Université de Lille Sciences et Technologies, Cité Scientifique, 59652 Villeneuve d'Ascq Cedex, France

*corresponding author: bernard.bonello@insp.jussieu.fr

**Abstract:**

We present the realization of the topological valley-protected zero-order antisymmetric ($A_0$) or symmetric ($S_0$) and zero-order shear-horizontal ($SH_0$) Lamb waves at different domain walls based on topologically distinct asymmetric double-sided pillared phononic crystals. The elastic periodic structures have either the triangular or the honeycomb symmetry and give rise to a double-negative branch in the dispersion curves. By artificially folding the doubly negative branch, a degenerate Dirac cone is achieved. Different polarization-dependent propagation along the same primary direction along the constituent branches are presented. Moreover, divergent polarization-dependent phenomena along different primary directions along a given branch are also reported. By imposing two large space-inversion symmetry (SIS) breaking perturbations the topological phase transition is obtained. We show that the Berry curvature becomes strongly anisotropic when the wave vector gets away from the valleys. Further, we demonstrate the unidirectional transport of $A_0$, $S_0$, and $SH_0$ Lamb waves at different domain walls in straight or Z-shape wave guides. In the large SIS breaking case, we show negligible reflection at the zigzag outlet of the straight wave guide and occurrence of weak inter-valley scattering at the bending corners of the Z-shape wave guide. For a larger strength of SIS breaking, the edge states are gapped and strong reflection at the zigzag outlet and bending corners is observed. The topological protection cannot be guaranteed any more in that case.

## I. INTRODUCTION

Topologically protected edge states provides an fascinating approach to manipulate the propagation of waves [1–10] since it allows for the one-way propagative edge states immune to defects and disorders at the domain walls between two topologically distinct configurations. The foundation of such a compelling phenomenon originates from the notion of topological phase which was first investigated in quantum systems [11–13]. Soon afterwards, its great potential for the unidirectional transport with negligible attenuation has been exploited for other types of waves, including acoustic and elastic waves, and several designs of phononic crystals have been proposed for this purpose [14–16].

Three kinds of topological phononic crystals have been reported, each of which exhibiting transport properties analogous to quantum Hall effect (QHE), quantum spin Hall effect (QSHE), and quantum valley Hall effect (QVHE) respectively. To mimic QHE, active components must be involved in order to break the time-reversal symmetry (TRS). In elastic systems this can be achieved with rotating gyroscopes and non-inertial platforms [2,17,18] or with flow circulators in acoustic systems [1,19,20] but at the price of great challenges in both the manufacturing and assembly. Emulating QSHE with TRS preserved only involves passive components, whereas this generally requires intricate designs to achieve a double degenerate Dirac cone [21–24]. As regards QVHE, it exploits the valley degree of freedom, which has been proved to be another controllable degree of freedom for an electron in a two-dimensional honeycomb lattice of graphene, and was quickly introduced in acoustic and elastic phononic crystals. Comparatively to the other two effects, it allows to significantly reduce the geometrical complexity and the topological phase transition can be simply achieved by breaking the space-inversion symmetry (SIS) in the unit cell. As a result, a great deal of configurations emulating QVHE have been proposed to theoretically and experimentally demonstrate the unidirectional transport of elastic waves both in discrete models [18,25–27] and in continuum structures [28–33]. Among these, the plate-type structures have generated much interests due to their advantages in controlling Lamb waves that are widely used in engineering applications, including nondestructive evaluation and structural health monitoring. For instance, the propagation of valley-protected antisymmetric Lamb waves in phononic plates filled with geometrical indentations [31] or triangular inclusions [33] has been reported. Another very promising platform is the pillared phononic crystal (PPnC) [14,34,35] built by regularly erecting cylinders on a thin plate [36,37]. These attached pillars act then as resonators which bending, compressional, and torsional modes can easily couple with the symmetric (S), the antisymmetric (A), or the shear-horizontal (SH) Lamb waves propagating in the plate respectively.

In this work, in analogy to QVHE, we have carried out the valley-protected $A_0$, $S_0$, and $SH_0$ Lamb waves at different domain walls constructed upon topologically distinct asymmetric double-sided PPnCs. Firstly, we describe a PPnC arranged in a triangular lattice, the Brillouin zone (BZ) of which features an isolated branch where both the effective mass density and the elastic modulus are negative. By offsetting the unit cell and rearranging the lattice in the honeycomb symmetry, the doubly negative branch is artificially folded, giving rise to a degenerate Dirac cone and to different polarization-dependent propagation along the same primary direction in the BZ, for each constituent branches. Moreover, we report on divergent polarization-dependent phenomena along the different primary directions on the same branch. Secondly, the topological phase transition is achieved by imposing a perturbation to the height of some pillars in the unit cell. Two SIS breaking situations with different strengths are considered. The Berry curvature in the vicinity of the valleys is numerically evaluated. This parameter becomes strongly anisotropic when the wave vector departs from the valleys. Thirdly, we investigate the unidirectional transport of Lamb waves along different domain walls either in a straight or in a Z-shaped wave guide. We show that the propagation of $SH_0$ Lamb wave is topologically protected at one domain wall where the propagation of $A_0$ ($S_0$) Lamb wave is forbidden because of the mismatch in the spatial parity. The opposite phenomenon is observed at the other domain wall. In the case of a moderate SIS breaking, the reflection at the zigzag outlet of the straight wave guide can be neglected and weak inter-valley scattering occurs at the bending corners of the Z-shape wave guide. For larger strength of SIS breaking, the edge states are gapped. A strong reflection occurs at the zigzag outlet of a straight wave guide and at the bending corners of a Z-shaped wave guide, which signifies the failure of the topological protection.

## II. CONSTRUCTING A DEGENERATE DIRAC CONE AND ITS POLARIZATION-DEPEDENT PROPAGATION

### A. Doubly negative branch in triangular lattice

In this section, we describe the procedure we used to construct a degenerate Dirac cone, which is a prerequisite for topological edge states to occur. It is based on the differentiated manner the eigenmodes of the pillar couple with the Lamb modes. Actually, in the low frequency regime, both $A_0$ and $S_0$ Lamb waves propagating in the plate can be altered by the bending and the compressional vibration of the pillars [38], whereas $SH_0$ mode, because of its in-plane polarization perpendicular to the propagation direction, can easily couple with the torsional resonance of the pillars [39]. Moreover, if the resonators in an asymmetric double-sided square lattice PPnC are thoughtfully designed, the bending, the compressional, and the torsional modes fall within the same frequency interval [39] and an isolated doubly negative branch occurs. The propagation along ΓX direction for a mode in this branch strongly depends on the polarization: it is allowed for an incident $SH_0$ Lamb wave, while it is not for $A_0$ or $S_0$ Lamb mode.

In analogy, we put forward an asymmetric double-sided PPnC arranged in a triangular lattice. The structure and the unit cell are displayed in Fig. 1(a). It features distinct pillars denoted as pillar A and pillar B concentrically connected to a thin plate. Both the plate and the pillars are made of steel whose Young's modulus, Poisson's ratio and mass density are $E = 200$GPa, $v = 0.3$ and $\rho = 7850$kg/m$^3$ respectively. The lattice constant and the thickness of the plate are $a = 231$μm and $e = 100$μm. Pillar A is designed to have both the bending and the compressional resonances within a common frequency interval where the effective mass density turns negative and the propagation of $A_0$ or $S_0$ modes is forbidden. The diameter and height of pillar A are $d_A = 120\mu m$ and $h_A = 268\mu m$. On the other hand, the dimensions of pillar B ($d_B = 140\mu m$ and $h_B = 160\mu m$) are optimized for the torsional resonance to occur in a narrow frequency interval inside this band gap. The propagation is allowed in this frequency interval, which shows that the doubly negative property is achieved therein. This could be formally demonstrated by calculating the band structure. The result, obtained by applying periodic conditions on the four lateral boundaries of the plate and solving the eigenvalue equations, is displayed in Fig. 1(b). As expected, a band gap opens up in between 4.176 and 4.643MHz and an isolated branch with a negative-slope (*i.e.* the group velocity is different from zero) takes place in between 4.408 and 4.486MHz. This branch is not present in the dispersion curves if one considers the single-sided PPnC constructed solely by pillar A or pillar B (not shown here). Therefore, at any frequency in between 4.408 and 4.486MHz, the system simultaneously exhibits negative effective mass density and shear modulus, whereas only the effective mass density is negative outside this band. The eigenmodes at points labelled C, D and E in Fig. 1(b) are displayed in Fig. 1(c). Clearly, the eigenmodes at points D and E are respectively the second-order bending resonance (relating to the components $\rho_{11}$ and $\rho_{22}$ of the effective mass density matrix) and the first-order compressional resonance (relating to $\rho_{33}$) of pillar A [40]. The eigenmode at point C is the first-order torsional resonance of pillar B and relates to the negative effective shear modulus. Actually, the weighting of the deformation around *z*-axis of pillar B, associated to this mode can be defined by:

$$\xi = \frac{\iiint_{\text{Pillar B}} (\text{curl } \mathbf{U})_z^2 \, dV}{\iiint_{\text{Pillar B}} (\text{curl } \mathbf{U}) \cdot (\text{curl } \mathbf{U}) \, dV},$$

where $\mathbf{U}$ and $V$ are the displacement field and the volume of the unit cell respectively. The result, depicted by the color scale in Fig. 1(b), bears out that the torsion around *z*-axis is the dominant deformation for the modes on the doubly negative branch, especially when the wave vector is close to a critical point in the BZ. As a consequence, only $SH_0$ Lamb waves may couple with a mode in the isolated branch. This is further confirmed by the spectra of the transmission of $A_0$, $S_0$, and $SH_0$ modes along $\Gamma K_1$ and $\Gamma M_1$, represented by the black, red and blue solid lines in Figs. 1(d) and 1(e) respectively: the doubly negative branch acts as a forbidden band for incident $A_0$ or $S_0$ Lamb waves, whereas an incident $SH_0$ Lamb wave propagates with negligible attenuation along both primary directions. This suggests that one may form a degenerate Dirac cone with such a doubly negative branch, and in turn obtain a structure suitable to support the topologically protected propagation of $SH_0$ Lamb waves.

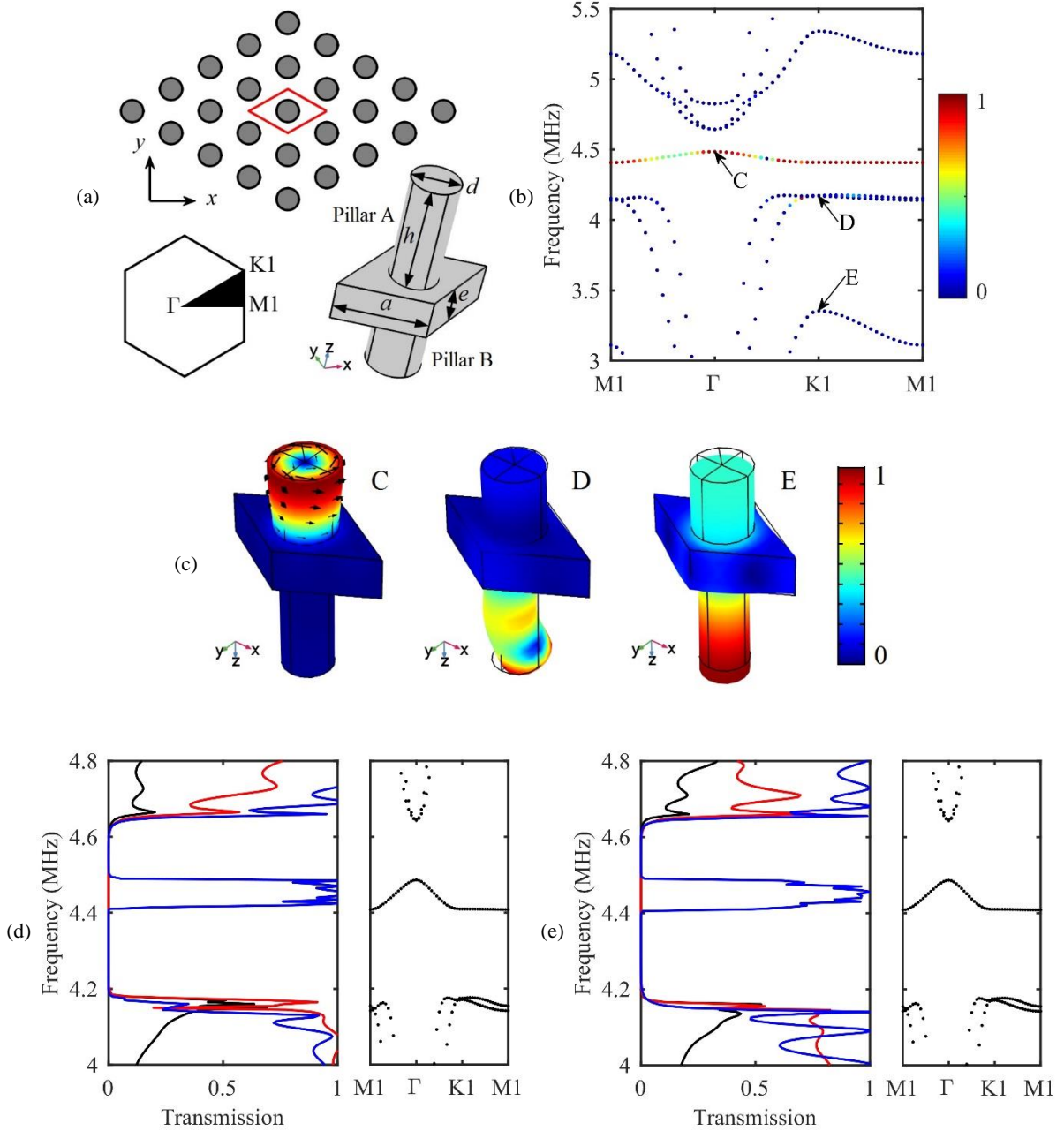

**Figure 1:** (a) Schematic of the asymmetric double-sided PPnC arranged in triangular lattice and (b) the corresponding band structure. (c) Eigenmodes at points labelled C, D and E in (b); the black arrows are for the deformation vectors. Transmission spectra of incident $A_0$ (black), $S_0$ (red) and $SH_0$ (blue) Lamb waves propagating along (d) $\Gamma K1$ and (e) $\Gamma M1$ directions.

## B. Artificial folding in honeycomb lattice

To construct a degenerate Dirac cone based on the aforementioned doubly negative branch, we considered the arrangement shown in Fig. 2(a). Two pairs of asymmetric double-sided pillars are assembled into a honeycomb unit cell, the distance between nearest neighbors being the same as for the triangular lattice depicted in Fig. 1(a). The lattice constant is set to $a = 400$μm. Actually, the honeycomb unit cell can be described as the merging of two triangular unit cells with an offset along ΓM2 direction between them. These two sub unit cells form equidistant Bragg plane only along ΓK2 direction, leading to a perfect artificial folding [41] and to a degenerate Dirac cone at point K2 of the BZ. Additionally, a thorough hole with diameter $d = 240$μm is drilled at each corner of the unit cell. These additional holes allow to soften the plate [42] without noticeable alteration of the band structure.

The corresponding band structure in the range $3 - 5.5$MHz is displayed in Fig. 2(b). Two eigenmodes at 4.318MHz create a degenerate Dirac cone at the corner K2. The group velocity is 56.14ms$^{-1}$ and a partial band gap opens up in between 4.302 and 4.344MHz along ΓM2 direction. Figure 2(c) depicts the deformations at points labelled as F, G, H1 and H2 in Fig. 2(b). At point F, both pillars BL and BR in the unit cell undergo in-phase torsional resonance. This, together with the group velocity which is negative around this point, suggests that the effective shear modulus negative is negative. As for the eigenmode at point G, both pillars BL and BR exhibit out-of-phase torsion which directly comes out from the band folding. Concerning the degenerate eigenmodes at points H1 and H2, torsional motions around $z$-axis in pillar BL as well as in pillar BR are predominant. It can be shown [18] that, once the degenerate Dirac cone is lifted by imposing SIS breaking perturbation, the new eigenmodes at the bounding are well approximated by a linear combination of the degenerate eigenmodes. It can be therefore speculated that the torsion resonances of both pillars BL and BR still play an important role in the perturbed configuration. The transmission spectra along ΓK2 for incident $A_0$ (black), $S_0$ (red) and $SH_0$ (blue) Lamb waves are shown in Fig. 2(d). The corresponding pattern is shown on the top. The excitation lays on the left of the supercell and two perfectly matched layers (PMLs) are placed at both ends to eliminate any reflected wave. The propagation of an incident $SH_0$ Lamb wave is allowed in the upper band (in between 4.318 and 4.353MHz, upper cyan region) but it is not for incident $A_0$ or $S_0$ Lamb waves. The opposite polarization-dependent propagation can be observed at frequency the lower band (in between 4.141 and 4.318MHz, lower cyan region). Moreover, in between 4.302 and 4.344MHz (magenta region), any Lamb mode can propagate, whatever the symmetry is, owing to the combination of in-phase and out-of-phase deformations. The situation is pretty different along ΓM2, as evidenced by Fig. 2(e). Clearly in that case, only $SH_0$ mode at a frequency either in the upper or in the lower band, can propagate. This can be recognized by considering the symmetry about $xz$-plane of the supercell drawn on top of each figure.

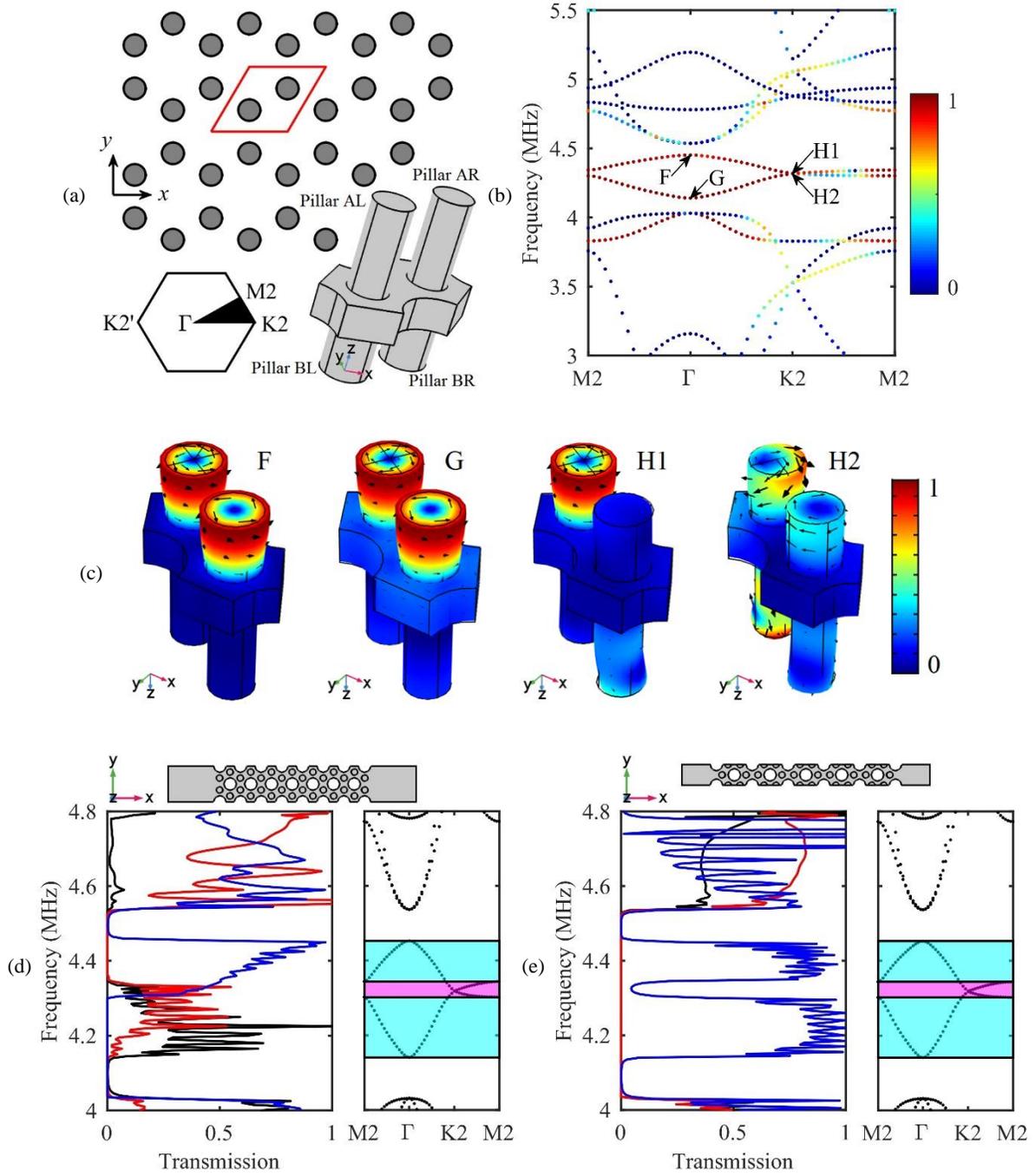

**Figure 2**: (a) Schematic of the asymmetric double-sided PPnC arranged in honeycomb lattice and (b) the corresponding band structure. (c) Eigenmodes at points labelled F, G, H1 and H2 in (b); the black arrows are for the deformation vectors. Transmission spectra of incident $A_0$ (black), $S_0$ (red) and $SH_0$ (blue) Lamb waves propagating along (d) ΓK2 and (e) ΓM2 directions.

## III. VALLEY-PROTECTED TOPOLOGICAL LAMB WAVES

### A. Topological phase transition

Mimicking QVHE requires that SIS perturbation is imposed to the honeycomb unit cell in order to lift the degenerate Dirac cone and to induce the topological phase transition. A non-zero valley Chern number can be obtained this way, along with the occurrence of a reopened band gap at both valleys. In practice, this can be done by varying either the height or the diameter of the asymmetric double-sided pillars. In doing so, the point group $C_{3v}$ is broken because of the violation of the mirror-symmetry about the mid plane of the unit cell, while the symmetry $C_3$ is preserved.

As discussed in the previous section, the torsional vibration of the lower pillars significantly contributes to the formation of the constituent branches of the degenerate Dirac cone. Therefore, we first consider the perturbation in the height of the lower pillars (pillars BL and BR) according to $h_{BL} = h_B + \Delta h$ and $h_{BR} = h_B - \Delta h$. The band structure of the perturbed PPnC with $\Delta h = 1\mu m$, hereafter referred as PPnC-I, is displayed in Fig. 3(a). As expected, the degenerate Dirac cone is lifted and an omnidirectional band gap ranging from 4.301 to 4.335MHz reopens. The vortex chirality at the valley K2 is revealed by the phase distribution of the out-of-plane displacement on the top surface of the plate, displayed in Fig. 3(b). In this figure, the top and bottom images represent the phase field at the higher (4.335MHz) and at the lower (4.301MHz) bounding frequencies respectively. In the top image, the phase field is centered on pillar AL and gradually decreases counterclockwise, while it keeps a constant value at the location where pillar AR is standing: this can be recognized as the valley pseudospin up state. In the bottom image, a uniformly distributed phase field is observed at the position of pillar AL, whereas it gradually decreases clockwise from the center of pillar AR, which corresponds to the valley pseudospin down state. It should be noted that these valley pseudospin states will be inverted at the valley K2' if considering TRS preservation in the system. Note also that each valley state can be selectively excited by imposing the proper chirality to fit with the pseudospin state of the desired valley [43–46]. The width of the topological band gap is fairly proportional to the magnitude of the height perturbation $\Delta h$. This may be checked in Fig. 3(c) where the pseudospin up and down states at valley K2 are drawn as red and black circles respectively. In the investigated range, the band gap firstly closes, reopens when $\Delta h$ crosses zero, whilst exchanging the frequency order of the up and down states which are signatures of the topological phase transition.

The configuration where the perturbation of the height $\Delta h = -1\mu m$, hereafter referred as PPnC-II, is exactly the space-inverted counterpart of PPnC-I. Obviously both configurations have the same band structure but nonetheless, they are absolutely distinct from the topological point of view. As a matter of fact, perturbing the height of the upper pillars (pillars AL and AR) according to $h_{AL} = h_A + \Delta h$ and $h_{AR} = h_A - \Delta h$ leads to the pseudospin up and down states displayed in Fig. 3(c) as black and red crosses. The comparison with the previous case clearly shows that the band gap is more sensitive to the height perturbation of the lower pillars.

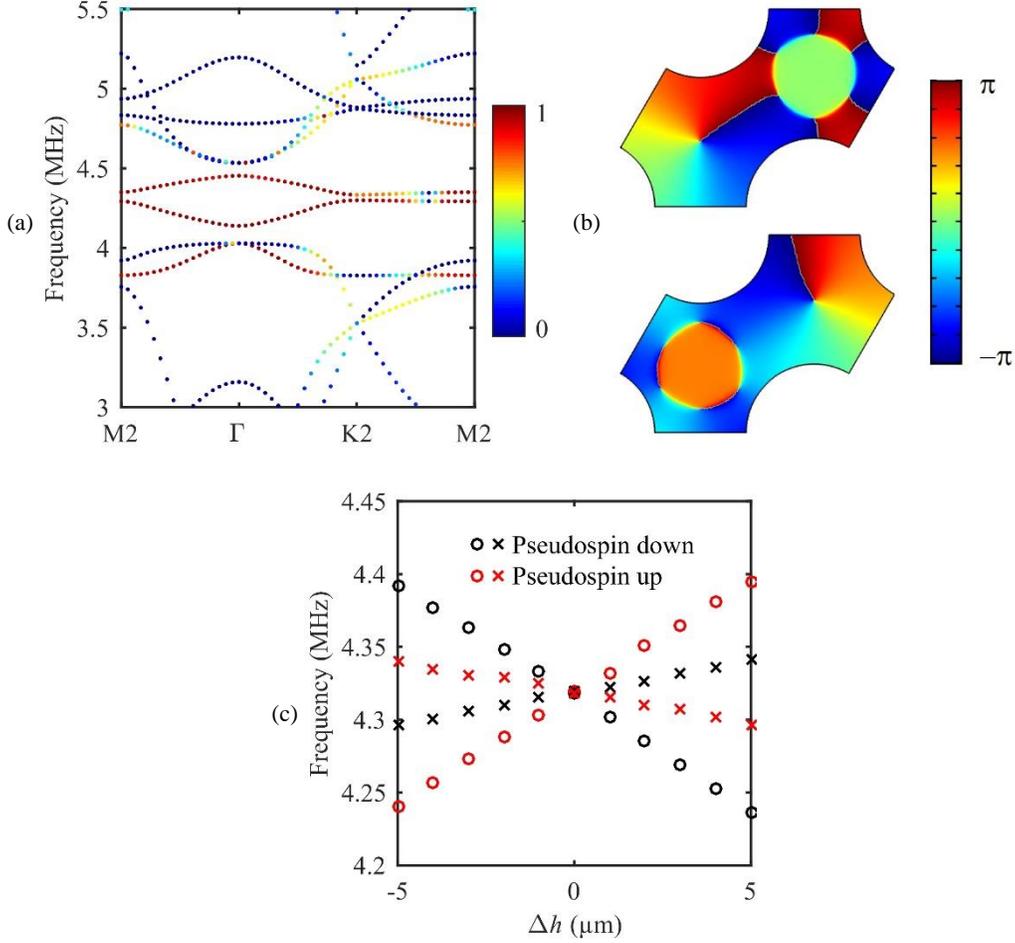

**Figure 3:** (a) Band structure of PPnC-I with the height perturbation $\Delta h = 1\,\mu m$. (b) Phase distribution of the out-of-plane displacement on the top surface of the plate at the higher (top panel) and lower (bottom panel) bounding of the lifted Dirac cone. (c) Evolution of the pseudospin up and down states at the valley K2 against the height perturbation $\Delta h$ in the lower (black and red circles) and upper (black and red crosses) pillars.

### B. Berry curvature and valley Chern number

In this subsection we calculate the integral over a portion of the BZ of the Berry curvature of PPnC-I, which is the invariant characterizing the topological nature of the structure after lifting the degenerate Dirac cone. The Berry curvature of the $n$th band at a given wave vector $\mathbf{k}$ can be calculated from:

$$\Omega_n(\mathbf{k}) = \nabla_{\mathbf{k}} \times \langle \mathbf{u}_n(\mathbf{k}) | i \nabla_{\mathbf{k}} | \mathbf{u}_n(\mathbf{k}) \rangle \cdot \hat{\mathbf{z}}, \tag{1}$$

where $\mathbf{u}_n(\mathbf{k})$ is the periodic part of the displacement field in the unit cell for the wave vector $\mathbf{k} = (k_x, k_y)$. Because of the TRS conservation, the topological invariant known as the Chern number calculated by integrating the Berry curvature over the whole BZ should be zero. Concerning SIS breaking in the perturbed configuration, the Berry curvature of the lower bounding of the lifted Dirac cone at the valley K2/K2' reads [32,33,47]:

$$\Omega(\mathbf{k}) = \pm \frac{m v_g^2}{2\left(|\delta \mathbf{k}|^2 v_g^2 + m^2\right)^{-2/3}}, \tag{2}$$

where $\delta \mathbf{k} = \mathbf{k} - \mathbf{k}_{K2/K2'}$ is the relative wave vector with respect to the valley K2/K2', $v_g$ is the group velocity at the degenerate Dirac cone; $m$ stands for the effective mass and is proportional to the frequency interval between the pseudospin up and down states. It is also directly related to the height perturbation $\Delta h$ and represents the strength of SIS breaking. In the small SIS breaking regime, the Berry curvature is strongly localized around the valleys K2/K2' and the valley Chern number $C_v$ quickly converges to a non-zero quantized value which can be theoretically derived as being [27,47]

$$C_v(\text{K2/K2'}) = \frac{1}{2\pi} \int \Omega(\mathbf{k}) d^2\mathbf{k} = \pm \frac{1}{2} \text{sgn}(m). \tag{3}$$

Equation (3) establishes that $C_v$ only depends on the sign of the effective mass $m$. On the other hand, Eq. (2) shows that the distribution of the Berry curvature broadens as the effective mass $m$ increases and interferences between two opposite valleys might occur. The assumption of a small SIS breaking is no longer valid and a large SIS breaking regime should be considered instead. In that case, when directly integrating Eq. (2) in the vicinity of the valleys [31] to calculate the valley Chern number, a huge deviation from the theoretical values ±1/2 might arise because of the destructive interference between two opposite valleys of the Berry curvature. Therefore, to identify the strength of SIS breaking in PPnC-I, we have computed the distribution of the Berry curvature around the valleys [33,47]. In Fig. 4(a), the black dotted line represents the numerical results along ΓK2 direction when the wave vector varies from $k_x = 2\pi/3a$ to $k_x = 10\pi/3a$. For comparison, the theoretical values predicted by Eq. (2) is displayed as a red solid line. The Both methods yield consistent results in between $k_x = 4\pi/3a$ and $k_x = 8\pi/3a$, i.e. along the high symmetry boundaries K2-M2-K2' of the BZ. From $k_x = 2\pi/3a$ to $k_x = 4\pi/3a$ the numerical results decrease more quickly than the theoretical estimate.

We have also computed the anisotropy around the valley K2. The result is depicted in Fig. 4(b) as black, red, and blue solid lines which represent the absolute values of the Berry curvature for the wave vector at a distance $|\delta \mathbf{k}| = 0.05|\Gamma \text{K2}|$, $0.1|\Gamma \text{K2}|$, and $0.2|\Gamma \text{K2}|$ away from the valley K2, respectively. The Berry curvature displays a circular shape when the wave vector is very close to the valley K2 that can be therefore considered as isotropic. However, it turns into a triangular shape as the distance |δk| increases up to 0.1|ΓK2| and even it changes to the clover-like shape when |δk| = 0.2|ΓK2| and exhibits then strong variations if the wave vector deviates from the high symmetry boundaries K2-M2-K2' in the BZ. This behavior relates to the steep dispersion curve around the lifted Dirac cone. The Berry curvature around the valley K2 when the components $k_x$ and $k_y$ vary in the intervals [π/a,2π/a] and [−√3π/3a ,√3π/3a] respectively, is displayed in Fig. 4(c). Integrating the Berry curvature over these intervals yields to a value for the valley Chern number significantly different from what is expected from theory since we found $C_v$= −0.22 instead of −1/2. This clearly indicates that the perturbation Δh = 1μm in PPnC-I corresponds to a large SIS breaking case. Generally, a larger SIS breaking case would occur if increasing the height perturbation. Figure 4(d) depicts both the numerical (black dotted line) and theoretical (red solid line) Berry curvature, when the height perturbation is set to Δh = 2μm. The profile of the Berry curvature is then much broader and the integral of the Berry curvature takes the value −0.08. In between $k_x$ = 4π/3a and $k_x$ = 8π/3a, the Berry curvature associated with the valley K2 and K2' tends to connect directly (see the slope) suggesting the occurrence of very strong inter-valley scattering.

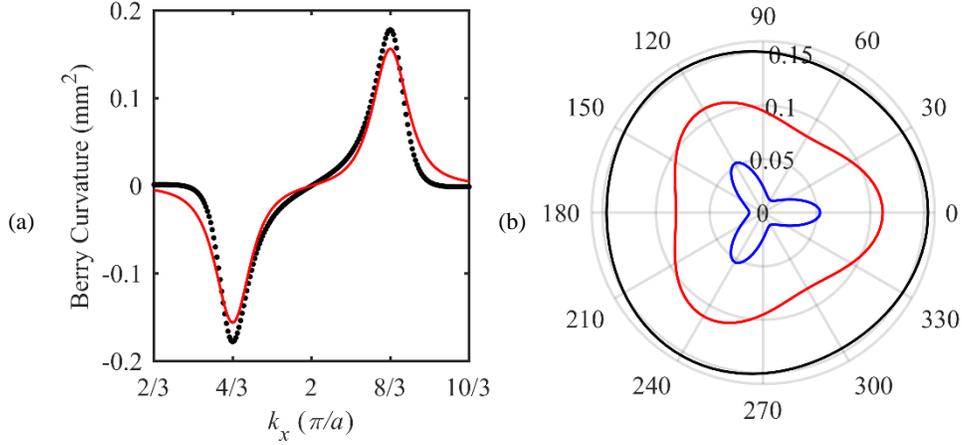

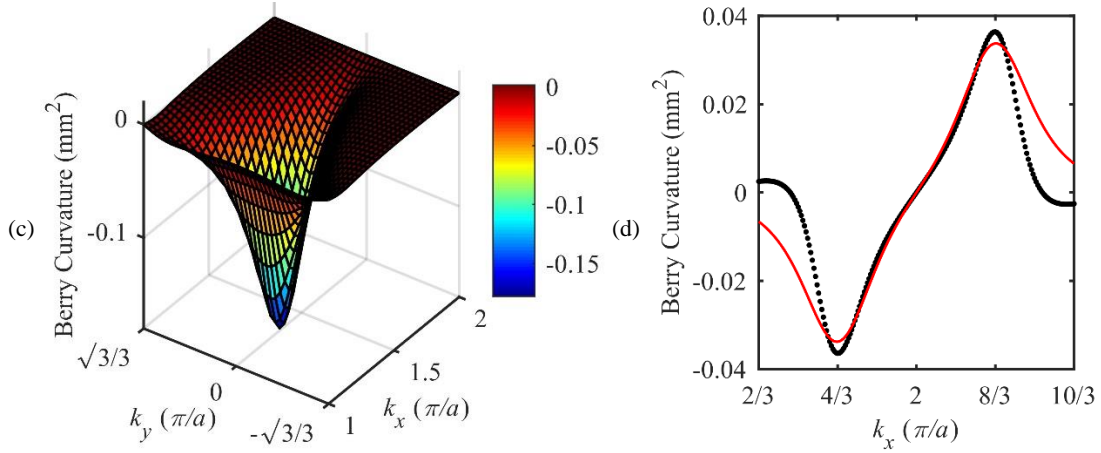

**Figure 4:** Numerical (black dotted line) and theoretical (red solid line) values of the Berry curvature along ΓK2 direction at the wave vectors varying from $k_x = 2\pi/3a$ to $k_x = 10\pi/3a$ and $k_y = 0$ when the height perturbation is set to be (a) $\Delta h = 1\mu m$ and (d) $\Delta h = 2\mu m$. (b) Absolute values of the Berry curvature at the wave vectors with a distance away from the valley K2 of $|\delta k| = 0.05|\Gamma K2|$ (black), $0.1|\Gamma K2|$ (red) and $0.2|\Gamma K2|$ (blue) respectively. (c) Distribution of the Berry curvature around the valley K2 at the wave vectors varying from $k_x = \pi/a$ to $k_x = 2\pi/a$ and from $k_y = -\sqrt{3}\pi/3a$ to $k_y = \sqrt{3}\pi/3a$.

### C. Valley-protected Lamb waves at the domain walls

The above analysis indicates the existence of the valley-protected edge states at the domain walls formed by the topologically distinct PPnC-I and PPnC-II. To verify this, we consider a three-layer ribbon supercell shown in Fig. 5(a), consisting of six unit cells of PPnC-I sandwiching eight unit cells of PPnC-II. Two zigzag domain walls are formed and zoomed in the side views, namely LDW and SDW, with the height of adjacent lower pillars at the interface expended and shortened by 1μm respectively. The dispersion curves of this ribbon supercell in the frequency interval ranging from 4.25 to 4.4MHz, are displayed in Fig. 5(b). The black dotted lines represent the bulk modes; the red and blue solid lines are the edge states occurring at LDW and SDW respectively. It should be noted that the projection of the valleys K2 and K2' onto these two zigzag domain walls are $k_x = -2\pi/3a$ and $k_x = 2\pi/3a$. At the valley K2, the group velocity of the edge state localized at LDW is negative as can be foreseen if considering the change of the valley Chern number from PPnC-I to PPnC-II, namely $C_v^{PPnC-I}(K2) - C_v^{PPnC-II}(K2) = -1$. In contrast, the group velocity is turning positive at the valley K2'. The eigenmodes at points M (4.316MHz) and N (4.323MHz) for the pillars on the rear face of the supercell are displayed in the left and right panels of Fig. 5(c), with the red arrows denoting the deformation vectors. Adjacent pillars on this face exhibit antisymmetric torsion about LDW interface while adjacent pillars on the upper face display antisymmetric bending (not shown here). In contrast, the torsion of the pillars on the rear face and the bending of the pillars on the upper face are symmetric about SDW interface.

It has been reported [3,48] that the antisymmetric edge state leads to a deaf band because of the mismatch in the spatial parity between the eigenmode and the deformation caused by the incident wave. We demonstrate in what follows that this edge state may be actually excited in our system if considering an incident $SH_0$ Lamb wave.

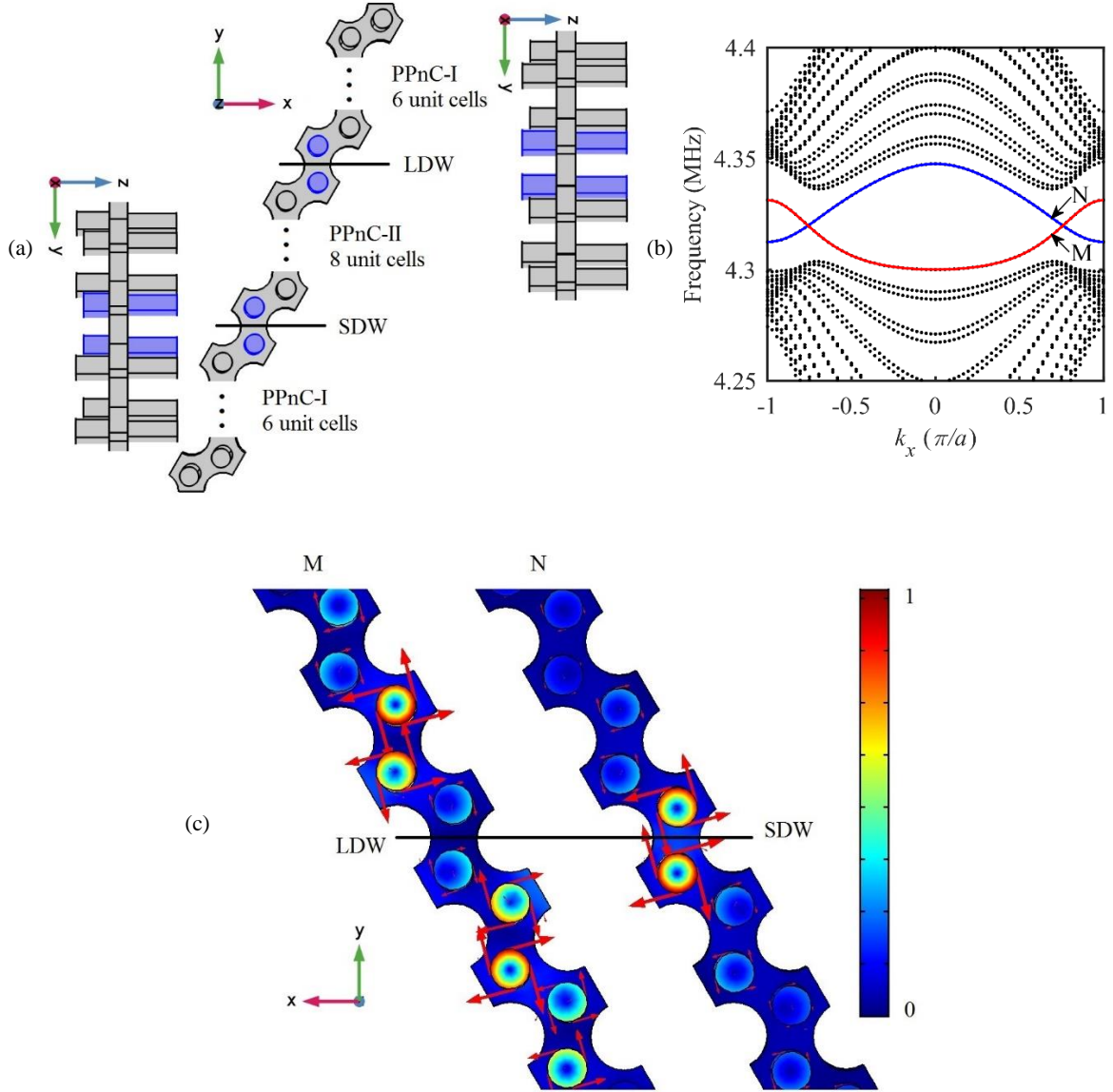

**Figure 5:** (a) Top view of the three-layer ribbon supercell constructed by placing six unit cells of PPnC-I at two ends and eight unit cells of PPnC-II in the middle and zoomed side views of two zigzag domain walls. (b) Dispersion curves of the ribbon supercell. (c) Bottom views of the eigenmodes at points labelled M and N in (b).

The straight wave guide featuring 24×20 unit cells shown in Fig. 6(a) was designed to investigate the unidirectional transport of the valley-protected edge states. The upper and lower domains are made of PPnC-I and PPnC-II that forms a LDW-type interface between them. PMLs enclose the structure to eliminate any reflected waves. In order to excite the K2'-polarized edge state [32], two elastic sources with a phase difference of $\pi/3$ and separated by a distance of $a$, are placed at the red point, where the edge state features antisymmetric deformation about the interface. An incident $SH_0$ Lamb wave at 4.314MHz is launched by applying $y$-axis polarized traction forces. The amplitude of the out-of-plane displacement on the top surface of the plate is plotted in Fig. 6(b) plots. Consistently with the positive group velocity at the valley K2', the wave propagates along positive $x$-axis only, toward the right zigzag termination where the K2'-polarized $SH_0$ Lamb wave gets both positively and negatively refracted. To quantitatively estimate the ability of the structure to prevent from backscattering, one may introduce the amplitude ratio between the out-of-plane displacement at the two ends, namely $\eta = A_{\text{Left}}/A_{\text{Right}}$. It is equal to 0.064 here. Although it slightly deviates from zero, we can still consider that we are in a large SIS breaking case and that the backscattering wave are actually suppressed at the right zigzag termination.

For comparison, Fig. 6(c) depicts the mapping of amplitude of the out-of-plane displacement when $z$-axis polarized forces are applied in order to excite $A_0$ Lamb wave at the same frequency. In that case, the elastic energy remains highly localized around the sources and cannot propagate along the domain wall because the generated field does not match the antisymmetric deformation of the edge state [see left panel of Fig. 5(c)]. It is therefore a forbidden band for an incident $A_0$ Lamb mode. The same conclusion holds if $S_0$ Lamb mode is excited.

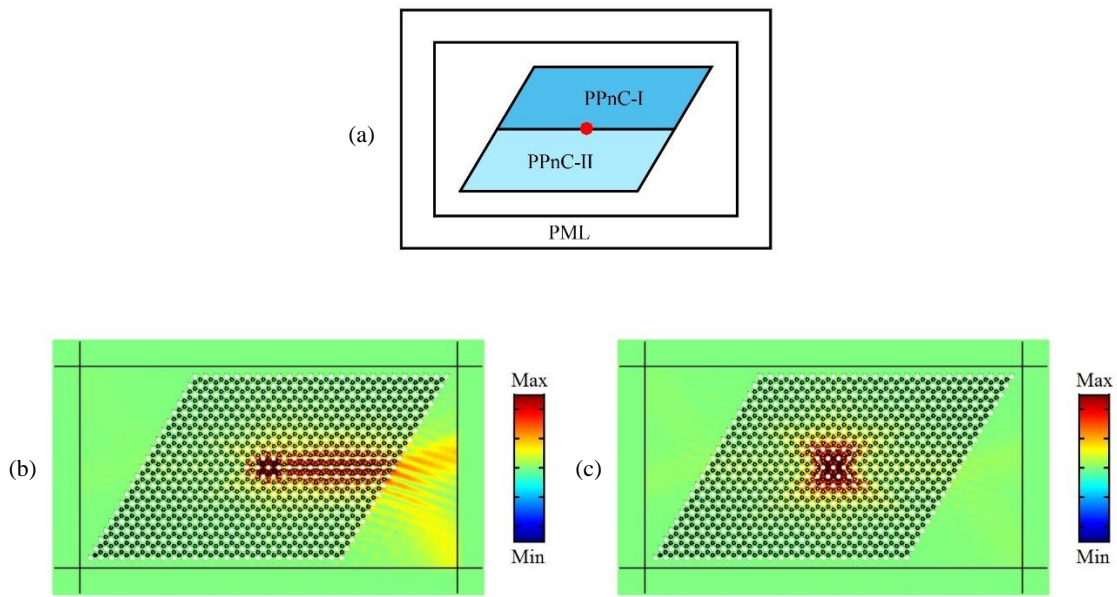

**Figure 6:** (a) Schematic of the straight wave guide featuring LDS-type interface constructed by placing PPnC-I and PPnC-II in the upper and lower domains respectively. The red point represents the position of two phase-matched sources. Plots of the amplitude of the out-of-plane displacement on the top surface of the plate under the excitation of the K2'-polarized (b) $SH_0$ and (c) $A_0$ Lamb waves at 4.314MHz.

Inverting the positions of PPnC-I and PPnC-II sketched in Fig. 6(a) allows to construct another straight wave guide featuring this time a SDW-type interface. Indeed, at this domain wall, the edge state displays a symmetric deformation about the interface. As previously did, K2'-polarized $SH_0$ and $A_0$ Lamb waves at frequency 4.325MHz are excited on the place marked by the red dot in Fig. 6(a). The amplitude of the out-of-plane displacement on the top surface of the plate are displayed in Figs. 7(a) and 7(b) respectively. whereas $SH_0$ Lamb wave remain localized around the sources, the excited $A_0$ Lamb wave propagate along negative *x*-axis. This behavior can be well understood when considering the spatial parity between the elastic field generated by the sources and the symmetric deformation of the edge state at this domain wall [see right panel of Fig. 5(c)]. In that case, only $A_0$ Lamb mode matches the required symmetric displacement field and the propagation along negative *x*-axis is consistent with the negative group velocity of the edge state at the valley K2'.

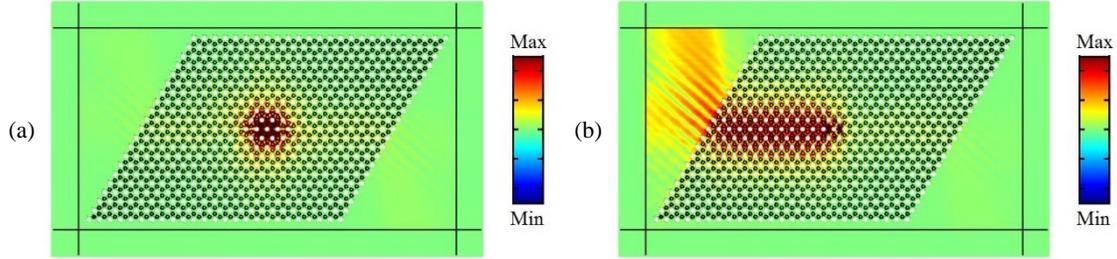

**Figure 7:** Plots of the amplitude of the out-of-plane displacement on the top surface of the plate under the excitation of the K2'-polarized (a) $SH_0$ and (c) $A_0$ Lamb waves at 4.325MHz. The straight wave guide is constructed by inverting the position of PPnC-I and PPnC-II shown in Fig. 6(a).

For the sake of completeness, we have investigated the propagation of the edge state in a Z-shape wave guide featuring two 60° sharp bending corners, as drawn in Fig. 8(a). The wave guide is created by placing PPnC-I and PPnC-II in the upper and lower domains respectively. According to the discussion above, only $SH_0$ Lamb wave can propagate along the domain wall. A right-going K2'-polarized $SH_0$ Lamb wave at 4.314MHz is launched from the red dot in Fig. 8(a). The resulting out-of-plane displacement field is displayed in Fig. 8(b). Refracted waves appear at the left outlet which indicates the occurrence of the inter-valley scattering of the edge state at the bending corners. Comparatively, the amplitude measured at the left outlet is much smaller than what is measured at the right outlet, suggesting weak inter-valley scattering. Actually, the amplitude ratio between the two outlets is 0.279, that is much larger than what is observed in the straight wave guide. This shows that, despite the weak inter-valley scattering occurring at the bending corners, most of the injected energy can propagate through the Z-shape wave guide within this large SIS breaking frame.

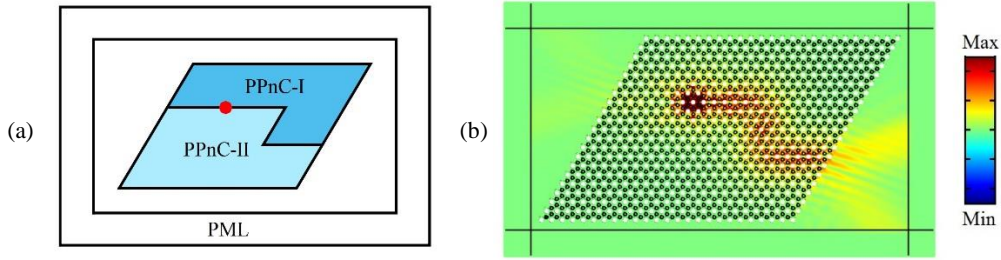

**Figure 8:** Schematic of the Z-shape wave guide constructed by placing PPnC-I and PPnC-II in the upper and lower domains respectively. The red point represents the position of two phase-matched sources. (b) Plot of the amplitude of the out-of-plane displacement on the top surface of the plate under the excitation of the K2'-polarized $SH_0$ Lamb wave at 4.314MHz in the Z-shape wave guide.

This last conclusion encourages to verify whether the pillared structure still allows for topological protection in an even larger SIS breaking situation. To this end, we have considered a ribbon supercell featuring the height perturbation $\Delta h = 2\mu m$, that gives rise to the dispersion curves displayed in Fig. 9(a) where the red and blue solid lines represent the edge state occurring at LDW and SDW respectively. These two edge states have similar profiles as the ones in Fig. 5(b), but are totally gapped. The topological protection has been examined by the propagation of the K2'-polarized $SH_0$ Lamb wave at 4.304MHz both in the straight and in the Z-shape wave guides. The results are shown in Figs. 9(b) and 9(c) respectively, where the waves reflected from the zigzag outlet and the bending corners can be clearly observed. The amplitude ratios are 0.228 for the straight wave guide and goes to 0.943 for the Z-shape wave guide. Therefore, it can be concluded that the topological protection of the edge state cannot be guaranteed any more in this larger SIS breaking case with gapped edge states.

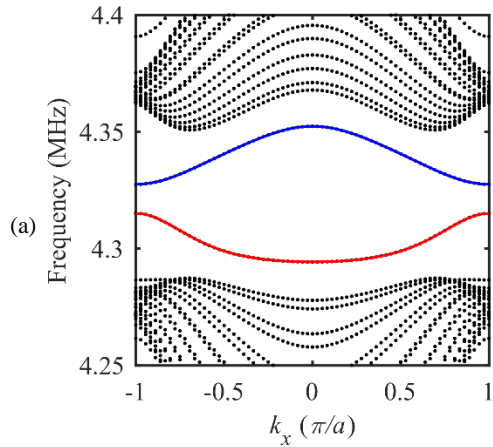

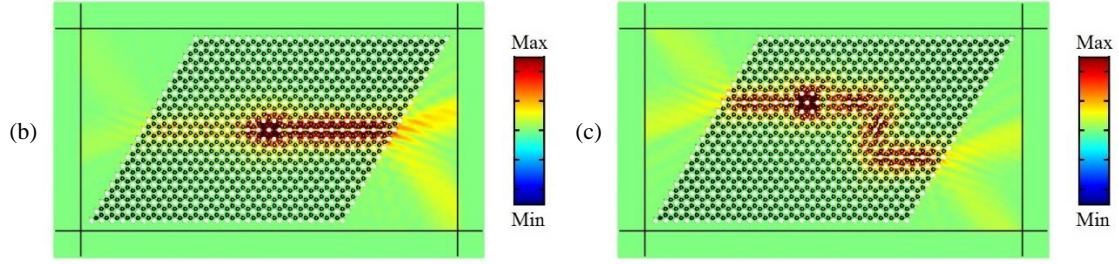

**Figure 9:** (a) Dispersion curves of the three-layer ribbon supercell with height perturbation $\Delta h = 2\mu m$. Plots the amplitude of the out-of-plane displacement on the top surface of the plate under the excitation of the K2'-polarized $SH_0$ Lamb wave at frequency 4.304MHz in both the (b) straight and (c) Z-shape wave guides.

## IV. Conclusion

In this work, the valley-protected topological propagation of $A_0$, $S_0$, and $SH_0$ Lamb waves at different domain walls constructed by topologically distinct asymmetric double-sided PPnCs is numerically demonstrated. A degenerate Dirac cone is achieved by artificially folding the doubly negative branch that occurs in the dispersion curves of both the triangular and the honeycomb lattices. At its constituent branches, different polarization-dependent propagation along the same primary direction of BZ are observed that are directly related to the in-phase and out-of-phase deformation in the honeycomb unit cell. Moreover, on a given branch, divergent polarization-dependent phenomena along different primary directions in the BZ are also reported. Afterwards, two large SIS breaking perturbations are imposed on the height of the lower pillars that features torsional motion to lift the degenerate Dirac cone, then realizing the topological phase transition. We observe that the Berry curvature becomes strongly anisotropic when the wave vector deviates from the valleys. Finally, the unidirectional transport of Lamb waves at different domain walls in the straight and Z-shape wave guides are discussed. The propagation of $SH_0$ Lamb wave is topologically protected at one domain wall where the propagation of $A_0$ or $S_0$ Lamb waves is forbidden because of the mismatch in the spatial parities. The contrary phenomena can be observed at the other domain wall. In the large SIS breaking case, the reflection at the zigzag outlet of the straight wave guide can be neglected and the weak inter-valley scattering occurs at the bending corners of the Z-shape wave guide. When the strength of SIS breaking becomes larger, the edge states are gapped and strong reflection at the zigzag outlet of the straight wave guide and the bending corners of the Z-shape wave guide is observed. The topological protection cannot be guaranteed any more in that case.